\documentclass[aps,prb,reprint,groupedaddress,showpacs,nofootinbib]{revtex4-2}
\usepackage{caption}
\usepackage{dcolumn}
\usepackage{bm}
\input{epsf}

\begin{document}

\title{Theory of electronic states in subwavelength two-dimensional nanostructures}
\author{L. Braginsky}
\affiliation{Rzhanov Institute of Semiconductor Physics, \\
Siberian Branch of the Russian Academy of Sciences, Novosibirsk 630090, Russia}
\affiliation{Novosibirsk State University, Novosibirsk 630090, Russia}
\author{M.~V.~Entin}
\affiliation{Rzhanov Institute of Semiconductor Physics, \\
Siberian Branch of the Russian Academy of Sciences, Novosibirsk 630090, Russia}

\begin{abstract}
Two-dimensional  nanosystems the characteristic sizes of which are less than the quasiparticle wavelength have been studied. This parameter allows replacement of the Shr\"odinger equation by the Laplace one. The latter permits us exact solution using the conformal mapping technique.
Bulged 1D quantum wires based on 2D system are considered.  The electron states in these systems have been studied.
The transmittance of the intersection between the narrow quantum strips has been studied. It is assumed that strip widths are less than the electron wavelength, so that they are the tunnel conductors. The transmittances of T-like and X-like wire crossings have been found.
\end{abstract}
\maketitle
\subsection*{Introduction}

The low-temperature conductance of small multiterminal quantum systems is described by the Landauer-Buttiker formula \cite{Land,Butt}:
$$J_i=\frac{e^2}{2\pi\hbar}\sum_{i\neq j}|T_{ij}|^2(V_i-V_j),$$
where $J_i$ and $V_i$ are the current and potential of the channel $i$, $T_{ij}$  is the transmittance amplitude.

The Landauer-Buttiker formula was used for description of the electron transport through the more complicated semiconductor structures: quantum dots \cite{Bandeira25,Fuhrer06}, rings \cite{Cesca23}, and carbon nanotubes \cite{Dai08}. The Aharonov-Bohm effect in quantum rings has been observed and thoroughly investigated in \cite{Alexeev13,Moldovan17,Huang25}.

Usually, the scattering matrix as well as the transmittance amplitude are considered for the propagating waves. In the present article, however, only the decaying waves are considered.   Although this situation seems  unusual, it is standard in various problems involving tunneling quantum junctions. This happens, for instance, when we consider the electron transport from the needle of the tunneling electron microscope.

Actually, various electronic tunneling devices also work on decaying states. Indeed, the conductance of the quantum rings used in the Aharonov-Bohm effect measurements is significantly less than the conductance quantum. In this case, the inputs and outputs to the ring should be tunneling. This is also the subject of our article.

The transmission/reflection amplitude matrices of the one-dimensional sections are multiplying. The same occurs with the cross junctions we are considering. Therefore, the quantities we have found here are important for coupling with the two-dimensional electron sea. The latter problem is not considered in the paper.

The reflection of the transverse quantization mode does not occur in the smooth adiabatic transition from a two-dimensional electron sea into a one-dimensional channel. If so, then the result for the transmission amplitude reduces to one that has been found in the paper. Of course, the interference at the arms of the Aharonov-Bohm interferometer could exist. Nevertheless, the tunneling through the intersection itself can be calculated using the Laplace equation.

A simplest multiterminal quantum nanosystem is a wire crossing, see  Fig.1, panels 1a) and 2a).
 Consider a   wire crossing, which  entrance and exit are small in comparison with the electron wave length.  The  Schr\"odinger equation in this case converts to the Laplace one. This permits its analytical solution. The purpose of the present paper is to find the transmittances of such  intersections.

In the recent paper \cite{we} we dealt with 2D quantum systems that have the sub-wavelength one-dimensional surface roughness. It was shown that such a rough boundary can be replaced by a flat one, however, shifted with regard to the mean surface position. This is correct if the roughnesses are small, but maybe not smooth.   Different physical problems at such boundary are reduced to this formulation, namely, the effective capacity of a flat capacitor,  resistivity of a conducting layer, reflection of the electromagnetic wave on the  metal surface,  laminar hydrodynamic flow in the rough  2D tube,  edge effects of the electron states in a  quantum layer and  wave resistance of a planar waveguide.  The solutions of these problems was found by the projection of them onto the exactly solvable by  conformal mapping 2D problems  of the Laplace equation  with the corresponding boundary.

In the present paper we consider the small subwavelength connections between two Fermi seas. We start with a simple 1D quantum wire connector. The problem is not trivial, if the wire contain a bulge. The localized state at this bulge essentially affects transmittance of the wire. We apply the conformal mapping that converts the bulged  wire into the quantum wire of constant width with the potential. Then we consider
 a small quantum wires intersection, which presents another exactly solvable problem.  Such  sub-wavelength systems occur, for example, as contacts of quantum rings utilized for the Aharonov-Bohm effect observation (see,i.g.\cite{aha}- \cite{fomin}  and references therein). These systems can be considered as the branching  1D wires. The calculation of the Aharonov-Bohm ring conductance can be reduced to  the entrance and exit branching transmittance.

The transmittances can be considered as the elements of $T$-matrix. However, unlike the usual situation when the incoming and outgoing solutions at infinity are presented by spatially oscillating plane waves, here the waves are exponentially decaying. The value of wave function at the  crossing is much less as compared with large distances. Hence, we should separate the total transmittance $T$, which includes the wire transmittances and the intrinsic transmittance $t$ of crossing caused by the crossing itself.

We  solve the Schr\"odinger equation $\Delta \psi+2mE\psi=0$ in the 2D wire crossing without potential in the wires, assuming the electron Fermi energy is essentially less than the quantization in the wires and their crossing. In these assumptions one can neglect the energy and the Schr\"odinger equation converts to the Laplace one.

 Usually, the transmittances are considered as given properties. They depend on the potential distribution within the conjuncture and are the subject of computer simulations \cite{kvon3}. Unlike these approaches, we consider an ideal potential-free system with boundaries. Such assumption permits analytic solution of the problem.

\subsection*{Electronic states in subwavelength two-dimensional nanostructures }
Consider a structure the characteristic scales of which are significantly smaller than the electron wavelength. This justifies the tunneling nature of particle motion. Such a situation corresponds to the near-field approximation in optics. From the scattering matrix consideration the problem corresponds to transitions between the evanescent electronic states instead of propagating waves. Although this approach is unusual, it is standard in various problems involving tunneling quantum point contacts.

A general feature of this approach is the replacement of the Schrödinger equation with the Laplace one in the region of interest. In a two-dimensional system this allows reducing the problem to an exactly solvable task of finding a conformal mapping of the considered region onto a simple one. Unlike the problem of electron motion in an external inhomogeneous potential such a problem has an exact analytical solution in many cases including the one-dimensional one.

\begin{figure}[h]

\centerline{\epsfysize=3cm\epsfbox{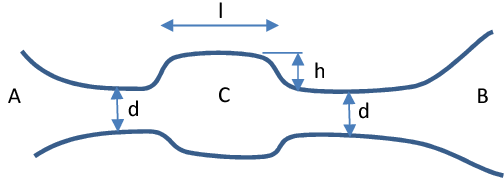}}

\caption {Uneven 1D channel between 2D seas A and B via two wires of widths $d$ and bulge C}\label{Fig1}

\end{figure}

\begin{figure}[h]

\centerline{\epsfysize=4cm\epsfbox{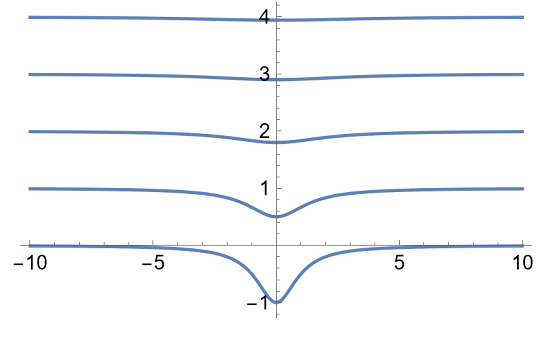}}

\caption{Lines of constant ${\rm Im}\,w(z)$ for conformal mapping  $w(z)=iFz + \frac{a}{z^2+b^2}$  with $a=10$, $b=1$, and $F=1$.  Any line can be considered as the wire boundary}\label{Fig2}

\end{figure}

\subsubsection*{One-dimensional quantum channel}

Note that although the transition from the Schrödinger equation to the Laplace equation is limited to low electron energies, this parameter range is highly relevant in the physics of low-dimensional systems.

We study the quasiparticle states with a simple parabolic spectrum in a two-dimensional nanostructure with an inflexible curved boundary. The quasiparticles can be either the charged carriers (electrons and holes) or excitons. Hereinafter we refer to them as electrons.  The motion of these quasiparticles obeys the Schrödinger equation. We consider the situation where the Schrödinger equation can be replaced by the Laplace one.  This is possible, if the electron energy is small in comparison with the size quantization energy.

Note that although the transition from the Schrödinger equation to the Laplace one is limited by the low electron energies, this parameter range is quite relevant in the physics of low-dimensional systems. In particular, this happens when low energy electrons tunnel from one two-dimensional Fermi sea to another one through a narrow channel.

Consider a general problem. The Laplace operator $\Delta$ can be written in complex form $\Delta=\partial_z\partial_{\tilde{z}}$, where $z=x+iy$. After the conform mapping  $z\rightarrow w(z)$ it becomes $\Delta=|\partial w/\partial z|^2\partial_z\partial_{\tilde{z}}$, and the 2D Schrödinger equation
\[
\Delta \psi+2m[E-V({\bm r})]\psi=0
\]
accepts the form
\[
\Delta_w \psi+2m[E-V({\bm r(w)})]\left|\frac{\partial z}{\partial w}\right|^2\psi=0.
\]
For $2m[E-V({\bm r(w)})]\left|\frac{\partial z}{\partial w}\right|^2\rightarrow 0$ this equation reduces to the Laplace one $\Delta_w\psi=0$. In the absence of the potential $V({\bm r})$ this means that either $2mEd^2\ll 1$ (here $d$  is the characteristic size of the system under consideration) or $|\partial z/\partial w|^2\ll 1$. The latter case corresponds to the slope or adiabatic system boundaries. It can be considered in the semiclassical approximation. We don’t consider it in this paper.

Consider now a quasi-1D system presented in Fig. The system contains 1D conducting channels of the same width d, which connect two 2D seas A and B through the bulge C. The size quantizing is not important in the seas where the Fermi energies are close to zero. On the contrary, the size quantizing in the channels and bulge exceeds the Fermi energy. The electron states in the bugle can be considered in semiclassical approximation, if  $h\ll l$. Then the problem reduces to 1D Schrödinger equation with the potential
\[
U(x)=\frac{\pi^2}{2md^2(x)}-\frac{\pi^2}{2md^2},
\]
where $d(x)$ is the channel width. If the Fermi energy in the seas are$E_F\ll {\pi^2}/{(2md^2)}$, then we have the tunnel channel. The local electron state in the bulge is ${\pi^2}/{(2md^2)}-E_0$, $E_0=(m/2) (\int_{-\infty}^{\infty}U(x)dx)^2$. Thus, the transmittance resonance occurs at $E=E_0$.

However, the semiclassical approximation is not applicable when $h\sim l$. If so, then the conformal mapping can be used. Let $w(z)$  maps the central part of the structure in Fig. to the plain region $0<y<d$  with the potential
\[
U(w)=\frac{\pi^2}{2md^2}\left(1-\left|\frac{\partial z}{\partial w}\right|^2\right).
\]

The resonance state is
\[
E_0=\frac{m}{2d^2}\left|\int_0^d\int_{-\infty}^{\infty}U(x+iy)\,dxdy\right|^2.
\]
This allows exact solution of the problem for any relation between $h$ and $l$.       It is apparent the solution yields the semiclassical limit at  $h\ll l$.

To be specific, consider the conform mapping
\[
w(z)=iFz+\frac{a^2}{z^2+b^2},
\]
where a,b, and F are reals. This yields the boundary $y=0$  with the bungle (Fig.). Assume another boundary to be plane. The mapping determines
\[
U(w)=\frac{\pi^2}{2md^2}\left(1-\left|F+\frac{2iaz}{(z^2+b^2)^2}\right|^2\right),
\]
where $z$  is the function of  $w$. We can replace the potential by  $\alpha\delta(w)$, if the longitudinal electron wavelength $1/p_F$  much exceeds the mean potential width. Here
\[
\alpha=\int U(w,\tilde{w})\theta[-i(w-\tilde{w})]\theta[2d+i(w-\tilde{w})]\,dwd\tilde{w}.
\]
Thus, the $\alpha$  value can be expressed via the problem geometry.

\subsection*{Wires. Solution at infinity}
In the wires of constant width $h$ the eigenfunctions of the Laplace equation $\Delta\psi=0$ are  $\psi_\pm\propto \exp(\pm\pi n x/h)\cos(\pi n y/hh)$, where $n$ is an integer.
In accordance with the causality on the long distance only decreasing towards the crossing solutions survive at incoming wires and, {\it vice versa}, at outgoing ones.

Consider the transmittance of a single wire. As the current is
\[
 J=\int dy i(\psi^*\partial_x\psi-\psi\partial_x\psi^*),
 \]
 the current-carrying solutions
should be composed from $\psi_\pm$  with complex coefficients.
The general solution in the n-th transversal mode reads $\psi_n(x)=\cos(\pi n y) (e^{-\pi nx}+r_ne^{\pi nx})$, where $r_n =i e^{-\pi nl}$ is chosen to fix the flow in the incoming wave 1 of  a wire with   length $l$.

In the cross geometry, one  can also consider the standing waves instead of decaying ones to obtain the transmission coefficients by separating decaying components
of the waves and treate their coefficients as transmission ones.

Owing to the wave function  decay, the long enough wires operate as  filters of the lowest transversal mode.
Considering the boundary conditions as an initial ones,    the long wires (independently of this condition) select  these  solutions.
Other modes occur are much  weaker.
This means, that one can set, for example, the constant function $\psi_\pm({\bf r})=c \exp(\pm\pi nx/h)$ across the wire at the entrances and project onto
the constant at the exits.

The crossings, considered below, contain long wires with exponentially small transmittances depending on their length. Their transmittance amplitudes are multiplied to the
intrinsic transmittance amplitudes of the crossings.

\subsection*{Conformal mapping}
The wave function obeys zero  conditions on the boundaries. This means that $\partial_{\bf t}\psi=0$, where ${\bf t}$ is the tangent to the boundary. The wave function should be complex for current-carrying states, therefore this condition is valid for real and imaginary parts of the wave function.

Besides, $\partial_{\bf n}\psi=0$ on the system symmetry line (here $x=0$ or $y=0$). Hence, $w(z)=w_1+iw_2=\partial_x\psi+i\partial_y\psi$ maps the quarter of Fig. b onto the $\pi/2$ angle between the rays $w_2=0,~w_1>0$, $w_1=0,~w_2>0$. The boundary converts to the positive abscissa, and the central line converts to the positive ordinate. Inside the source and the drain  $\psi= a \exp(-\pi n x/2h)\cos(\pi n y/2h)+b\exp(\pi n x/2h)\cos(\pi n y/2h)$. Here $n$ numerates the mode of transversal quantization in the wire.

In the cross-system b), there are independent transmittance amplitudes $t_{1,n_1, 2,n_2}=t_{2,n_2,3,n_3}=t_{3,n_3,4,n_4}=t_{4,n_4,1,n_1}$, $t_{1,n_1,3,n_3}$ and $t_{2,n_2,4,n_4}$. Besides, within the Laplace model, $t_{ij}=t_{ji}$. t
To find transmittance amplitude$t_{ij}$, one should block all wires $k\neq i,~k\neq j$ by additional borders at the infinities of these wires. In this case, the system with many wires converts to a two-exit system.

\subsection*{Connection of quantum wires intersections transmittance  with conductances of   multi-entrance electric circuites}

The problem of the Schr\"odinger equation with zero energy

is dual to the Laplace equation $\Delta\phi=0$ for the potential $\phi$ of the macroscopic current flow.  Their boundary conditions read:

$$\partial_{\bf n}\phi=0~~\mbox{Laplace}, ~~\psi=0~~\mbox{Schr\"odinger}. $$

These conditions should be supplemented by conditions at sources-drains

$$\phi=\phi_i~~\mbox{Laplace},~~ \partial_{\bf n}\psi=ik\psi~~\mbox{Schr\"odinger}. $$

The last condition follows from the conjunction with 2D seas. The potentials $\phi$ and $\psi$ obey the Couchy-Riemann conditions, so that the quantity $\omega=\phi+i\psi$ becomes the analytical function of  variable $z=x+iy$.

That means that the quantum circuit transmittance amplitudes $t_{ij}$  are  connected with $G_{ij}$ of similar electric circuit expressing the currents in wires via the potentials  $V_i$ in them. The potentials is assumed to be constant across the wire, the total current in a wire is found by integration of the local current densities $j_i=\sigma \partial_{\bf t}\phi$ across the wire. The transmittances are described by fixing the wave function amplitude  $a_i$  in a single entrance  and mode and fiding the wave function amplitudes  $a_j$  at exits. The absence of reflected waves  is assumed due to adiabatic conjunctions.

 The ratio of these amplitudes $a_j$ with account for the wire widths and electron translational momentum yields the crossing transmittances, $t_{ij}=\frac{a_j}{a_i}\sqrt{\frac{h_jp_j}{h_ip_i}}$, where $h_i$ is the wire $i$ width. This allows you to connect $t_{ij}$ and $G_{ij}$.

\hspace{2cm}

\begin{figure}[h]

\centerline{\epsfysize=8cm\epsfbox{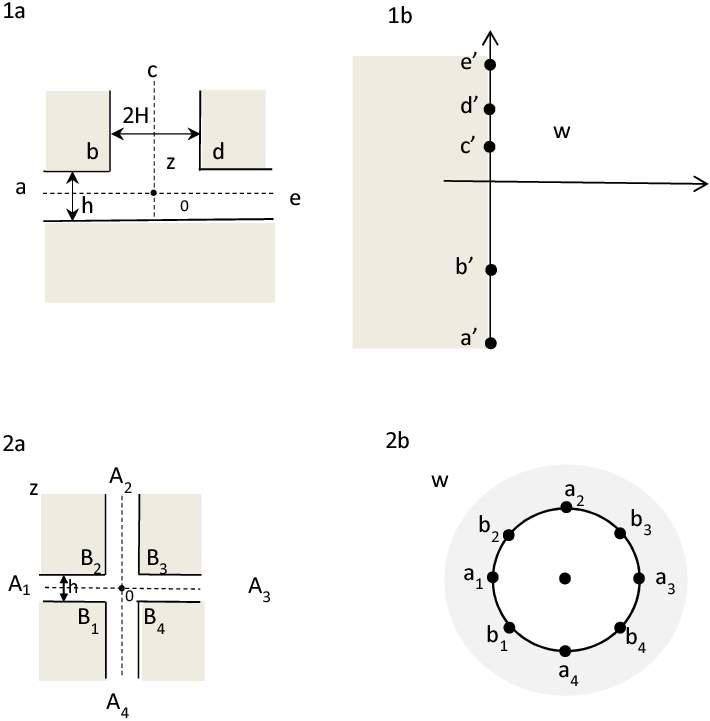}}

\caption{1) Sketch of the crossing between $ 3$ wires $i$, connecting 2D seas. 2) X-crossing. Panels a and b refer to  $z$- and $w$ planes, correspondingly.   }\label{Fig3}

\end{figure}

\begin{figure}[h]

\centerline{\epsfbox{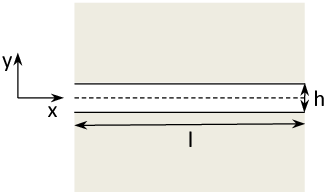}}

\caption{ Sketch of a single quantum wire.  }\label{Fig4}

\end{figure}

Now consider the conductances of the electrical circuit. The linearity of the problem together with the Kirhgoff laws yield s equations for

potentials $\phi_i$ and currents $j_i$ in the n-entrance circuit

\[
j_i=\sum_jG_{ij}\phi_j,~~\sum_jG_{ij}=0,~~G_{ij}=G_{ji}.
\]

Unlike the quantum problems, the boundary conditions

in this case are zero the normal derivatives of the potential. As a result, the potential in wires no more depends  across the wire

and  depends not exponentially, but linearly along the wire.

\subsection*{T-form crossing}
To be clear, we count the lengths of wires counted from the crossing of the central lines of the wires. In such case, the intrinsic crossing transmittance amplitude is the ratio of extrinsic transmittance amplitude
and transmittance amplitude of incoming/outgoing wires.

Consider the  case of a T-form wires crossing. The conformal mapping of the right half-plane $z$ onto this system in the $w$ plane (Fig.1) is given by the function

\begin{equation}\label{11112}
w(z)=\frac{h}{\pi}\left( \ln\frac{s\sqrt{z^2-a^2}+iz}{s\sqrt{z^2-a^2}-iz}+is\ln\frac{\sqrt{z^2-a^2}+iz}{\sqrt{z^2-a^2}-iz} \right)\end{equation}
where $s=H/h$, $a=\sqrt{s^2+1}$, all functions are analytically continued from the positive semi-abscissa and the cuts  are chosen  along the negative semi-abscissa.

The points $A_1,~A_2,~A_3$ in the $w$ plane and the corresponding points $a_1,~a_2,~a_3$ in the $z$ plain serve as sources-and-drains, with the exponential wave in the $w$ plane or multipoles located in the $z$ plane border. Substitution of the inverse function $z(W)$ into the multipole potential $\partial^n_z\log z$ gives the needed  solution in the $w$ plane.

\subsection*{X-form crossing}
Here we consider the X-form wire crossing. It can be treated as a limit of the square-shape star with rays going to infinity.  Using the system symmetry  (see \cite{Lavrentiev}) we find the conformal mapping of the circle onto the crossing
$$w=C\int_{-1}^z\frac{(1-z^4)^{3/4}}{1+z^4}dz.
$$
The constant $C$ is found from the equation
$w(1)-w(0)=h/\sqrt{2},$ that yields
$$\frac{C}{2^{5/4}}\int_0^{e^{i\pi/4}}\frac{dt}{(t+1)(t-1)^{3/4}}=he^{i\pi/4}/\sqrt{2}.$$
The points $a_k=e^{i\pi (2k+1)/4}$, where $k$ are integers in the $z$ plane correspond, to the wires in the $w$ plain. The points $b_k=e^{i\pi k/2}$ are images of rectangular angles. The zero potential on the border in the $w$  plane conserves in the $z$ plane. The source basic decaying wave in the $w$ plane converts into the dipole potentials $c_k Re(1/(z-a_k))$ in the $z$ plane. The coefficients $c_k$ determine the transmission amplitudes. Expressing $z$ via $w$ we find the wave function everywhere in the crossing area.

\subsection*{Discussion and conclusions}

We studied electron states in sub-wavelength 2D nanostructures. The small size of these systems allows reduction of the Schr\"odinger equation to the Laplace one, which permits using the powerful conformal mapping methods to solve it. We assume the wires to be narrower than the Fermi wavelength, so that the wave functions are decaying along the wires.

The quantum wires with bulges were considered. Our consideration concerns shapes that go beyond the limits of the semiclassical approximation. Besides, we  found the transmittances of small crossings of wires in the 2D system.   The T- and X-like crossing were considered. The problem is actual for small wire crossings. In particular, it  appears in the problems of Aharonov-Bohm effect in quantum rings. The knowledge of the transmission amplitudes permits modeling of  the Aharonov-Bohm effect amplitude.

Our results disregard the internal wire potential.
It should be also emphasized that the same method  of utilizing the Laplace equation instead of the Schr\"odinger one is applied to the transmittances of other small branchy 2D nanostructures.
For example,  consider the quantum intersections of $N$ narrow wires with rectangular walls connecting $i$ 2D seas, $i\leq N$.  An entire internal part of circuit is supposed to have a size less than the electron wavelength.  In this case electrons tunnel along the wires with wave functions $\propto e^{\pm \kappa_0y} \cos(\kappa_0y)$, where $\kappa_0= \pi y/2d_0$, $x$ and $y$ are coordinates along and across wires of widths $d_0$.


\begin{thebibliography}{8}

\bibitem{Land}     Landauer, R. (1957). "Spatial Variation of Currents and Fields Due to Localized Scatterers in Metallic Conduction". IBM Journal of Research and Development. 1 (3): 223-231. doi:10.1147/rd.13.0223

\bibitem{Butt}  Buttiker, M. (1990). ''Quantized Transmission of a Saddle-Point Constriction''. Phys. Rev. B 41 (11):7906-7909.  Bibcode:1990PhRvB..41.7906B. doi:10.1103/PhysRevB.41.7906. PMID 9993100

\bibitem{Bandeira25} N. S. Bandeira, Andrey Chaves, L. V. de Castro, R. N. Costa Filho, M. Mirzakhani, F. M. Peeters, and D. R. da Costa
Phys. Rev. B 111, 125409 (2025) - Published 13 March, 2025
\bibitem{Fuhrer06} A. Fuhrer, P. Brusheim, T. Ihn, M. Sigrist, K. Ensslin, W. Wegscheider, and M. Bichler
Phys. Rev. B 73, 205326 (2006) - Published 12 May, 2006

\bibitem{Cesca23} Joshua Cesca and Cédric Simenel,
 Phys. Rev. C 108, 054307 (2023) - Published 15 November, 2023


\bibitem{Dai08}W. Dai,  B. Wang, Th. Koschny, and C. M. Soukoulis
Phys. Rev. B 78, 073109 – Published 28 August, 2008
DOI: https://doi.org/10.1103/PhysRevB.78.073109

\bibitem{Alexeev13} A. M. Alexeev, I. A. Shelykh, and M. E. Portnoi
Phys. Rev. B 88, 085429 (2013) - Published 26 August, 2013

\bibitem{Moldovan17} L. L. Li, D. Moldovan, P. Vasilopoulos, and F. M. Peeters
Phys. Rev. B 95, 205426 (2017) - Published 22 May, 2017

\bibitem{Huang25} L. Huang, G. Wei, and A.R. Champagne
Phys. Rev. Applied 23, 014030 (2025) - Published 15 January, 2025


\bibitem{we}  L. S. Braginsky, M. V. Entin, JETP {\bf 168}, 765 (2025).

\bibitem{aha}van Oudenaarden, A., Devoret, M., Nazarov, Y. et al. Magneto-electric Aharonov-Bohm effect in metal rings. Nature 391, 768-770 (1998). https://doi.org/10.1038/3508

\bibitem{kvon1}     Gusev, G. M., Kvon, Z. D., Litvin, L. V., Nastaushev, Y. V., Kalagin, A. K., and Toropov, A. I.  Aharonov-Bohm oscillations in a 2D electron gas with a periodic lattice of scatterers. JETP letters, 55(2), 123 (1992).

\bibitem{kvon2}   Gusev, G. M., Kvon, Z. D., Shegai, O. A., Mikhailov, N. N., and Dvoretsky, S. A.  Aharonov Bohm effect in 2D topological insulator. Solid State Communications, 205, 4-8(2015).


\bibitem{kvon3}Tkachenko, V. A. E., Kvon, Z. D., Tkachenko, O. A., Baksheev, D. G., Estibals, O., and Portal, J. C. Coulomb blockade in a lateral triangular small quantum dot. Journal of Experimental and Theoretical Physics Letters, 76(12), 720-723 (2002).

\bibitem{kach1}    Dmitriev, A. P., Gornyi, I. V., Kachorovskii, V. Y., and Polyakov, D. G.  Aharonov-Bohm Conductance through a Single-Channel Quantum Ring: Persistent-Current Blockade and Zero-Mode Dephasing. Physical review letters, 105(3), 036402(2010)

 \bibitem{fomin} Physics of quantum rings. Second edition, ed. by Vladimir M. Fomin, ISSN 1434-4904 ISSN 2197-7127 (electronic) NanoScience and Technology ISBN 978-3-319-95158-4 ISBN 978-3-319-95159-1 (eBook) https://doi.org/10.1007/978-3-319-95159-1 Library of Congress Control Number: 2018947471. Springer International Publishing AG, part of Springer Nature Heidelberg (2018).

\bibitem{Lavrentiev}  M. A.  Lavrentiev and B. V. Shabat. "Methods of complex function theory." Moscow: Nauka (1987). 206.

\end{thebibliography}
\end{document}